# CuPc molecules adsorbed on Au(110)-(1x2): growth morphology and evolution of valence band states

*Fabrizio Evangelista[1,*], A. Ruocco[1], Valdis Corradini[2,3], M.P.Donzello[4],*
*Carlo Mariani[5,3] and Maria Grazia Betti[5]*

[1] Dipartimento di Fisica and Unità INFM , Università Roma Tre

Via Vasca Navale 84,I-00146, Roma

[2] Dipartimento di Fisica and Unità INFM, Università di Modena e Reggio Emilia

Via G. Campi 213/A, I-41100, Modena

[3] INFM National Center on nanoStructures and bioSystems at Surfaces (S^3)

Via G. Campi 213/A, I-41100, Modena

[4]Dipartimento di Chimica, Università di Roma "La Sapienza"

P.le A. Moro 2, I-00185 Roma

[5] Dipartimento di Fisica and Unità INFM, Università di Roma "La Sapienza"

P.le A. Moro 2, I-00185 Roma

[*]Corresponding author: *Tel. +39 06 55177221; Fax: +39 06 5579303;*

*E-mail: evangelista@fis.uniroma3.it*

We present the growth morphology, the long range ordering, and the evolution of the valence band electronic states of ultra-thin films of copper phthalocyanine (CuPc) deposited on the Au(110)-(1x2) reconstructed surface, as a function of the organic molecule coverage. The Low Energy Electron Diffraction (LEED) patterns present a (5x3) reconstruction from the early adsorption stages. High-Resolution UV photoelectron spectroscopy (HR-UPS) data show the disappearance of the Au surface states related to the (1x2) reconstruction, and the presence of new electronic features related to the molecule-substrate interaction and to the CuPc molecular states. The CuPc highest occupied molecular orbital (HOMO) gradually emerges in the valence band, while the interface electronic states are quenched, upon increasing the coverage.

Keywords: Low Energy Electron Diffraction (LEED), Ultraviolet Photoelectron Spectroscopy  (UPS), Growth, Interface states, Surface reconstruction

**Introduction**

Metal-phthalocyanine (MPc) films deposited on inorganic substrates, usually grow in columnar stacks, where the column/column interaction is very weak. For this reason the film conductivity is observed mainly along the stacking axis, parallel to the column direction, where the π-π interaction of the conjugated orbitals takes place [1]. These MPc films can exhibit different growth modes, characterized by different angles between the normal to the surface and the stacking axis. A strong influence of the substrate on the growth morphology of MPc thin films is well established. In particular, both the chemical nature and the surface geometry of the substrate are important parameters [2, 3]. It is well known that a flat-lying deposition is the most probable configuration for the first deposited layer [4] in presence of strong molecule/substrate interaction [5], while, when layered materials (i. e. mica or graphite) are employed as substrates, the CuPc molecules weakly interact, favouring a self-assembling behavior, and giving rise to columnar stacks with a different orientation angle [6]. It is worth noting that there are few studies on the early stages of adsorption of CuPc films on metallic substrates, and that the influence of the first deposited layer on the successive layers is far to be completely understood.

Furthermore, organic-inorganic heterostructures have become a very interesting research subject, because of the possibility of exploiting both the optical sensitivity of an organic material and the transport properties of an inorganic material [7]. In these systems, at the nanoscale range, even interfacial properties become important [8].

In this work, we study the early stages of CuPc adsorption on the Au(110)-(1x2) surface, in order to investigate the long range ordering, the electronic interaction between the molecule and the substrate, and how the molecule-surface interaction affects the electronic properties of the CuPc films. The Au(110) surface exhibits a very stable (1x2) missing-row reconstruction [9, 10] which has been recently used as a template to investigate the growth of organic molecules like thiophene and pentacene [11, 12]. After the first CuPc deposition, the Au(110) substrate evolves in a (5x3) reconstruction, with the formation of a long-range ordered two-dimensional (2D) Pc structural phase, as deduced by the LEED patterns. The valence band feature of this superstructure presents molecular-substrate interface states and CuPc molecular states. In order to discriminate the Pc molecular states from the interface states, the Pc induced electronic features have been compared with those of a thick film of Pc obtained after low temperature deposition on the same substrate. The interface electronic structure of the 2D ordered Pc layer can be mainly related to the x3 reconstruction of the Au(110) surface.

**Experimental**

The LEED and the HR-UPS measurements were performed at the LOTUS surface physics laboratory. The HR-UPS spectra were collected by means of a high angular and energy resolution Scienta SES-200 hemispherical analyser. The apparatus includes two experimental ultra-high-vacuum (UHV) chambers separated by a UHV gate valve: the main chamber hosts the high resolution electron spectrometer, the LEED apparatus and the helium discharge source (HeI$_\alpha$ and HeII$_\alpha$ photons, h$\nu$ = 21.218 eV and 40.814 eV, respectively), while the second UHV chamber is devoted to the preparation and cleaning procedures for the Au surface and for the CuPc film deposition. The Au sample is mounted on a four degrees-of-freedom manipulator, with 0.5 mm and 0.5 deg accuracy in the translation and angular coordinates, respectively. This configuration allows varying the sample temperature from 100 K to 1000 K.

The analyzer was used in the angle integrated mode, collecting ±6$^o$ around the normal to the sample surface, along the [1$\underline{1}$0] direction. Angle-integrated UPS has widely employed to minimize photoelectron diffraction effects [13] in the study of the surface and interface states as a function of the adsorbate coverage. A good compromise between signal-to-noise ratio, good energy resolution, and acquisition time, was obtained using 2 eV pass energy and 0.8 mm slits in the angular integrated mode. Under these conditions, the energy resolution is 16 meV, as determined on the Fermi level (E$_F$) of the Au specimen. Calibration of the binding energy scale was carried out using the Au Fermi edge, measured at 16.890 eV of kinetic energy.

A sharp long-range ordered (1x2) reconstruction of the clean Au(110) surface was obtained by means of a double-step sputtering-annealing treatment: in the first cycle the Au surface is Ar-bombarded at the sputtering energy of 1 keV and the sample is heated up to 725 K, while in the second cycle the ion bombarding energy is lowered to 0.5 keV with an annealing temperature of 530 K [14].

CuPc was purchased from Aldrich Chemical (97% dye content) and purified by means of two sublimation cycles. The low vapour pressure and the high thermal stability [1] make MPc a suitable choice for depositing thin films in UHV conditions. Ultrathin films of CuPc have been prepared by evaporation from a crucible, using an e-beam evaporator [15], at a very low deposition rate (less than 1 Å/min). The film thickness and the evaporation rate were estimated by means of a quartz microbalance. The nominal thickness is referred to an equivalent uniform bulk-like layer with a density of 1.62 g/cm$^3$. In order to obtain ultrathin films of increasing thickness, CuPc molecules were progressively evaporated on the substrate kept at room temperature (RT), up to saturation coverage. The saturation coverage was defined when no further evolution in the photoemission spectra is detectable (see discussion in the next section). A thicker layer was obtained by depositing CuPc on the interface obtained at room temperature, by keeping the sample at 100 K (low temperature, LT). In order to minimize flux intensity fluctuations in the UPS spectra, the He discharge lamp was maintained in the same work conditions, between two consecutive CuPc depositions in the preparation chamber.

**Results and discussion**

High energy-resolution photoemission measurements performed at room temperature, for CuPc films of different thickness, are reported in Fig.1. After a nominal coverage of 9 Å, the Energy Distribution Curve (EDC) spectra do not show any significant evolution even at much higher coverages (about 40 Å, not shown here). The expected escape depth at 15 eV of kinetic energy (close to the top of the valence band) is of about 15 Å [16], well beyond the thickness value for which the valence band appears unchanged. Nonetheless, the CuPc molecular bulk configuration is not reached: in fact, the Au structures are still visible, the highest occupied molecular orbital (HOMO) state, expected at ~1.3-1.5 eV binding energy (BE), is a weak feature, and the molecular levels at higher binding energy [17] are not detectable. The deposition of pentacene on the Au(110) surface at RT, gives rise to a similar behaviour of EDC as a function of the coverage [12, 18]; from a structural point of view LEED pattern shows a first 2D ordered layer, while Atomic Force microscopy (AFM) images present the nucleation of three-dimensional (3D) linear stripes on top of the first layer [18]. Consistent with the Pentacene/Au interface results, the EDC data reported in this work can not exclude the formation of a 3D CuPc phase, leaving wide free patches of the underlying 2D ordered layer, though they are compatible as well with the existence of a saturation coverage, that is a sticking coefficient equal to zero.

A selection of the valence band EDC of the CuPc/Au(110) system at different coverages, obtained with the substrate kept at room temperature, is reported in Fig. 2. The higher coverages reported in the figure (lower spectra) are obtained enhancing the sticking probability, by keeping the sample at 100 K. All the CuPc/Au UPS spectra, but the clean Au(110)-1x2, belong to the same set of consecutive measurements.. The clean gold valence band structure exhibits the d-band states in the energy region between 2-6 eV BE, while a lower cross section emission is detectable at lower binding energy, where the density of states has a dominant s-character [19]. The surface states (S) arising from the missing-row (1x2) reconstruction are detected in the spectra at 3.01 eV and 1.97 eV of BE, superimposed on the bulk density of states. Theoretical calculations assign a surface resonance at 2 eV BE [9], while a surface state at higher binding energy with a dominant d-character is predicted at 4 eV at the $\overline{\Gamma}$ point of the surface Brillouin zone.

After the first Pc deposition (2 Å) the Au(110) surface states are strongly quenched, while three new structures (I) located at 4.34, 3.14 and 2.67 eV of BE emerge, and the structure at about 6 eV of binding energy (present in the clean Au spectrum) is enhanced. Upon increasing the CuPc film thickness, a gradual intensity decrease of the I structures and the change in the spectral lineshape of the peak at 6 eV, is noticeable. The measurements at higher coverages, namely 14 Å and 20 Å, obtained with the substrate kept at LT, show two new features at 3.63 and 1.35 eV of binding energy (M), while the interface (I) structures are completely quenched. It is also interesting to follow the evolution of a less intense electronic state, that appears as a shoulder centered at about 4.9 eV of BE at the lowest coverage. This structure, as the CuPc thickness grows, moves towards higher binding energies up to 5.0 eV where a well defined structure is visible at 20 Å.

Following the CuPc/Au(110) electronic state evolution, in the low coverage range we notice that the I structures (with increasing intensity up to a coverage of 4 Å) are quenched as the film thickness increases, so that we can ascribe to the formation of interface states. It is worth noting that analogous new structures at similar binding energies have been also detected for pentacene ($C_{22}H_{14}$) adsorption on the same substrate [12]. The shoulder at about 4.9 eV could also be interpreted as an interface state, as it fades upon increasing the CuPc coverage, while the further spectral evolution towards higher binding energies could be explained with an overlapping of this state with a molecular state, detectable at higher coverage. Regarding the structure at about 6 eV, the change in the lineshape and the intensity behaviour cannot allow an unequivocal attribution, because of the coexistence of a Au state and of an interface level, arising after the first deposition and quenched upon increasing the CuPc film thickness. The M peaks, clearly detectable for the spectra collected at low temperature, correspond to the CuPc highest occupied molecular orbitals [17] and, in particular, the state at lower binding energy is commonly known as the HOMO state.

As far as the HOMO state is concerned, it is interesting to analyze in more detail its evolution as a function of coverage. In fig. 3 we report an enlargement of the HR-UPS spectra collected at room and low temperature, in the low binding energy region. A very weak structure is detectable at about 1.2 eV BE on the clean Au surface, while a more intense double peak develops in the same energy region after CuPc deposition, showing no significant lineshape evolution, apart from the different intensity increasing of the two components, till the saturation value is reached at RT. In particular, for a film thickness of 2 Å, only the peak at lower binding energy is detectable, while, starting from a coverage of 4 Å, the second peak centered at higher binding energy is visible, becoming dominant upon increasing the CuPc film thickness. The two structures lie at about 1.15 and 1.42 eV BE for a film thickness of 10 Å. The measurements taken on the molecular layers grown at LT show an asymmetric broad structure, centered at 1.33 eV.

Theoretical calculations suggest the presence of a Au surface state located at about 1 eV below the Fermi level [9], which might explain the weak structure at 1.2 eV BE, although we cannot exclude a valence band peak excited by the HeI$_\beta$ satellite line (23.085 eV photon energy, and 1/50 intensity with respect to the peak excited by HeI$_\alpha$). The HOMO double-peak structure persists at high coverage, so that we can reasonably assign these peaks to bulk-like occupied molecular levels. A double absorption structure in the same energy range has been very recently observed by energy loss spectroscopy for a CuPc-Au(100) interface prepared at RT [20]: this structure has been addressed to as a CuPc bulk feature of the ordered molecular solid, in agreement with previous optical absorption data [21].

The relative intensity behavior of the two components with increasing the film thickness, suggests a role of the interface in the first coverage steps. The peaks broadening and the consequent presence of an unresolved double-peak structure at 1.33 eV can be explained with the formation of a disordered multi-layer phase at LT, neglecting the actual interface region.

The electronic valence band structure of the CuPc-Au(110) system brings to light proper molecule-substrate interaction states at the actual interface, as well as molecular levels. The line shape evolution of the molecular levels at low binding energy as a function of the substrate temperature, seem to suggest the presence of a disordered phase at low deposition temperature.

The LEED patterns of the clean Au(110)-(1x2) reconstructed surface detected at the electron energy of 42 eV (A), and of the 3 Å-CuPc/Au(110) system taken at 45 eV (B), are reported in Fig. 4. In the same figure, the LEED intensity distributions along the [001] and the [$1\bar{1}0$] directions, for the clean Au and the CuPc/Au interface, are also reported. For the clean surface a well defined (1x2) reconstruction has been obtained, with sharp doubled spots along the [001] direction, revealing the presence of missing rows along the [$1\bar{1}0$] direction [9,10]. The structural ordering of the CuPc/Au(110) system shown in Fig. 4B, appears dramatically different. In particular, taking as a reference the (1x1) surface unit cell in the reciprocal space, a x3 reconstruction is observed along the [001] direction, while in the perpendicular direction a x5 periodicity is visible with broadened and elongated spots along the [001] direction. The (5x3) superstructure is observed for an extended primary electron beam energy range (15-60 eV). Besides, the intensity distribution along the [001] shows a clear x3 reconstruction for integer value of h, being h the component of the surface reciprocal vector in the [$1\bar{1}0$] direction (see fig. 4D). For fractional values of h, instead of sharp spots, we observe a diffuse intensity.

In order to get information on the structure of the interface it is important to compare the dimension of the molecule with the periodicities observed in the LEED patterns. CuPc is a planar, substantially square molecule with a side of 13.8 Å [1]. The x5 reconstruction corresponds, in the real space, to a periodicity of 5x2.88 Å=14.4 Å, that well matches the CuPc molecular size, while the x3 reconstruction corresponds to a periodicity of 12.2 Å. We can assume an alignment of the CuPc molecules forming long-range ordered molecular chains along the [$1\bar{1}0$] direction, while the molecular dimensions are not compatible with a flat-lying commensurate structure along the [001] direction. In particular, based on the estimated figures, we can expect a tilt angle at least of 30°. From the intensities distribution reported in fig. 4D, we observe that fractional spots due to the clean Au (1x2) reconstruction are completely missing in the CuPc/Au system, thus implying that the clean Au (1x2) surface reconstruction is deeply modified. Despite the observed x3 reconstruction along the [001] direction could be due to CuPc disposition with respect to unrelaxed substrate, we believe this is not the case. In fact we observe along the [001] direction a 3 fold reconstruction only for integer value of h at all the beam energies. As a consequence we suggest the CuPc molecules induce a profound rearrangement of the gold substrate, that undergoes the observed x3 reconstruction. There are several experimental and theoretical works accounting for a Au(110)-(1x3) reconstruction induced by atomic or molecular adsorption on the Au(110) surface. In particular, such a gold reconstruction has been observed upon alkali metals deposition [22] or even upon organic molecules adsorption [11, 12, 18]. Moreover, from theoretical calculations, it comes out that the (1x3) reconstructions is energetically close to the (1x2) [23], and both are highly energetically favored with respect to the (1x1) surface symmetry.

While a commensurate growth is evident along the $[1\bar{1}0]$ direction, present experimental data do not permit to conclude which is the orientation of molecular planes with respect to the substrate. On the other hand, a recent study of the CuPc/Au(111) interface has shown a flat-lying molecular growth with respect to the substrate [24]; in the hypothesis of x3 reconstruction of Au surface and considering that the walls of the channels are oriented along the (111) surface, it is plausible that the molecule has its plane parallel to the micro-facets of the trough. However, further investigations with suitable structural techniques are clear necessary in order to completely resolve the structure of the interface and, in particular, to determine the orientation of the molecule with respect to the underlying gold surface.

**Conclusions**

Copper-Phtalocyanine deposition on the Au(110)-(1x2) surface produces distinct interface and molecular states in the valence band region, while the surface states related to the Au (1x2) reconstruction disappear, as measured by high energy resolution UPS. In particular, the molecule/Au interface states emerge for a nominal coverage of about 2 Å, and are progressively reduced at higher coverage, while the molecular levels grow. A distinct (5x3) LEED pattern is observed from the lowest coverages, marking a long range ordered 2D CuPc phase. From LEED images, we can suggest a molecular alignment along the $[1\bar{1}0]$ corresponding to the rows of the Au substrate. In the [001] direction we observe the disappearance of the clean Au x2 reconstruction and the appearance of a x3 periodicity: we believe that this symmetry is due to the formation of Au channels parallel to the $[1\bar{1}0]$ direction.

**Acknowledgements**

Fruitful experimental collaboration with Claudia Menozzi is gratefully acknowledged.

**References**

[1] M. J. Stillman, T. Nyokong in *"Phthalocyanine - Properties and Applications"*, C. C. Leznoff, A. B. P. Lever Eds., VCH Publ., New York, vol. 1, 1989.

[2] J. J. Cox, S. M. Bayliss, T. S. Jones, *Surf. Sci.*, 425 (1999) 326

[3] J. J. Cox, S. M. Bayliss, T. S. Jones, *Surf. Sci.*, 433 (1999) 152

[4] G. Dufour, C. Poncey, F. Rochet, H. Roulet, M. Sacchi, M. De Santis, M. De Crescenzi, *Surf. Sci.*, 319 (1994) 251

[5] A. Ruocco, M. P. Donzello, F. Evangelista, G. Stefani, *Phys. Rev. B*, in press

[6] T. Shimada, K. Hamaguchi, A. Koma, F. S. Ohuchi, *Appl. Phys. Lett.*, 72 (1998) 1869

[7] J. Takada, H. Awaji, M. Koshioka, A. Nakajima, W. A. Nevin, *Appl. Phys. Lett.*, 61 (1992) 2184

[8] J. Takada, H. Awaji, M. Koshioka, M. Imanishi, N. Fukada, W. A. Nevin, J. *Appl. Phys.*, 75 (1994) 4055

[9] C. H. Xu, K. M. Ho, K.P. Bohnen, *Phys. Rev. B*, 39 (1989) 5599

[10] K.-M. Ho and K. P. Bohnen, *Phys. Rev. Lett.*, 59 (1987) 1833

[11] S. Prato, L. Floreano, D. Cvetko, V. De Renzi, A. Morgante, S. Modesti, F. Biscarini, R. Zamboni, C. Taliani, *J. Phys. Chem. B*, 103 (1999) 7788

[12] V. Corradini, C. Menozzi, M. Cavallini, F. Biscarini, M. G. Betti, C. Mariani, *21st European Conference on Surface Science (ECOSS 21), 2002; Surf. Sci.,* in press

[13] C. Solterbeck, W. Schattke, J.-W. Zahlmann-Nowitzki, K.-U. Gawlik, L. Kipp, M. Skibowski, C. S. Fadley, M. A. Van Hove, Phys. Rev. Lett., 79 (1997) 4681

[14] A. Hoss and M. Nold, P. von Blanckenhagen, O. Meyer, *Phys. Rev. B*, 45 (1992) 8714

[15] R. Verucchi, S. Nannarone, *Rev. of Scient. Instr.*, 71 (2000) 3444

[16] S.Hufner, "*Photoelectron Spectroscopy*", Springer, (1995)

[17] I.G. Hill, A. Kahn, *J. Appl. Phys.*, 86 (1999) 2116

[18] C. Menozzi, V. Corradini, M. Cavallini, F. Biscarini, M. G. Betti, C. Mariani, E-MRS Spring Meeting *2002; Thin Solid Films,* in press

[19] D. A. Shirley, *Phys. Rev. B*, 5 (1972) 4709

[20] J.M. Auerhammer, M. Knupfer, H. Peisert, J. Fink, *Surf. Sci.* 506 (2002) 333.

[21]  E. A. Lucia, F. D. Verderame, *J. Chem. Phys.*,48,  (1968), 2674.

[22] P. Haberle, P. Fenter, T. Gustafsson, *Phys. Rev. B*, 39 (1989) 5810

[23] M. Garofalo, E. Tosatti, F. Ercolessi, *Surf. Sci.*, 188 (1987) 321

[24] I. Chizhov, G. Scoles, A. Kahn, *Langmuir*, 16 (2000) 4358

**Figure captions**

Fig. 1: Angle integrated HR-UPS valence band spectra (hν=21.218 eV) as a function of the CuPc film thickness, grown on the Au(110) substrate at room temperature.

Fig. 2: Angular integrated HR-UPS spectra of CuPc/Au(110) at different coverages on the substrate kept at RT. The spectrum at 10 Å corresponds to the saturation coverage; the lower spectra are taken on layers grown on the substrate kept at low temperature (T=100 K).

Fig. 3: Enlargement of the HR-UPS spectra reported in fig.2 in the HOMO energy region. Spectra at the highest coverages are referred to measurements on the system grown at low temperature (LT). The clean Au spectrum is relative to a different set of measurements.

Fig. 4: LEED pattern of the clean Au(110) (1x2) surface taken at 42 eV(A) and of the CuPC/Au(110) (5x3) interface taken at 45 eV (B). The rectangle, in the clean gold LEED pattern, represent the (1x1) unit cell of the unreconstructed surface; the main crystallographic axis are also reported.

Right side: Intensity distribution of the diffraction signal along the main surface direction: [1$\bar{1}$0] (C) and [001] (D); for each direction are reported, in the same figure, the diffraction intensities of the clean gold surface and the diffraction intensities of the CuPc/Au(110) interface.

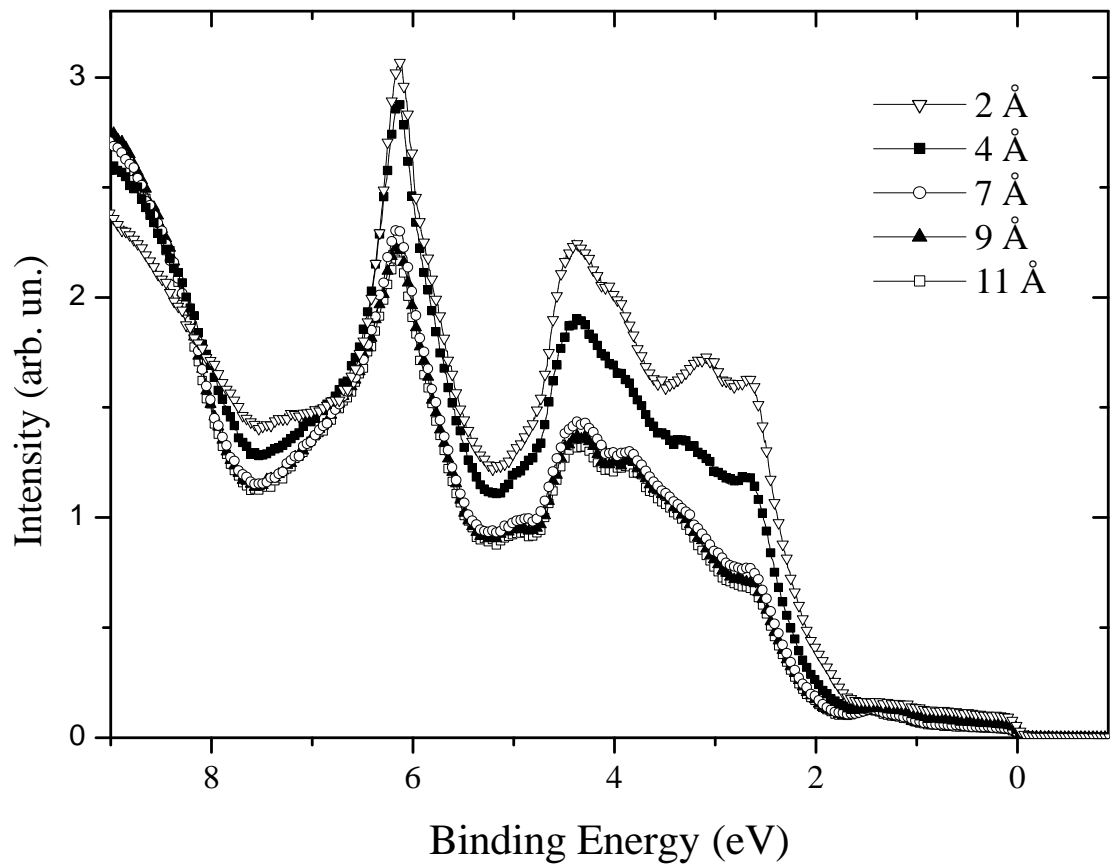

Fig.1 Evangelista et al.

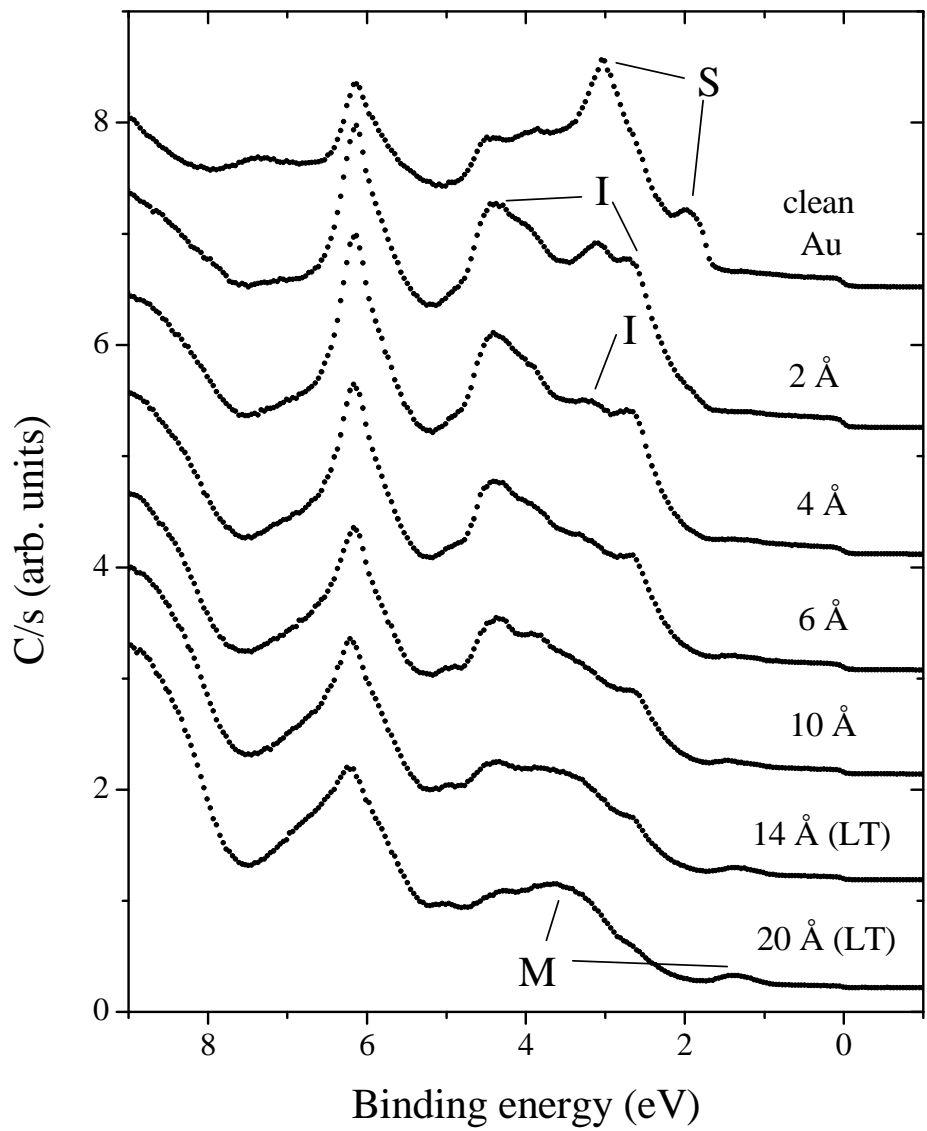

Fig.2 Evangelista et al.

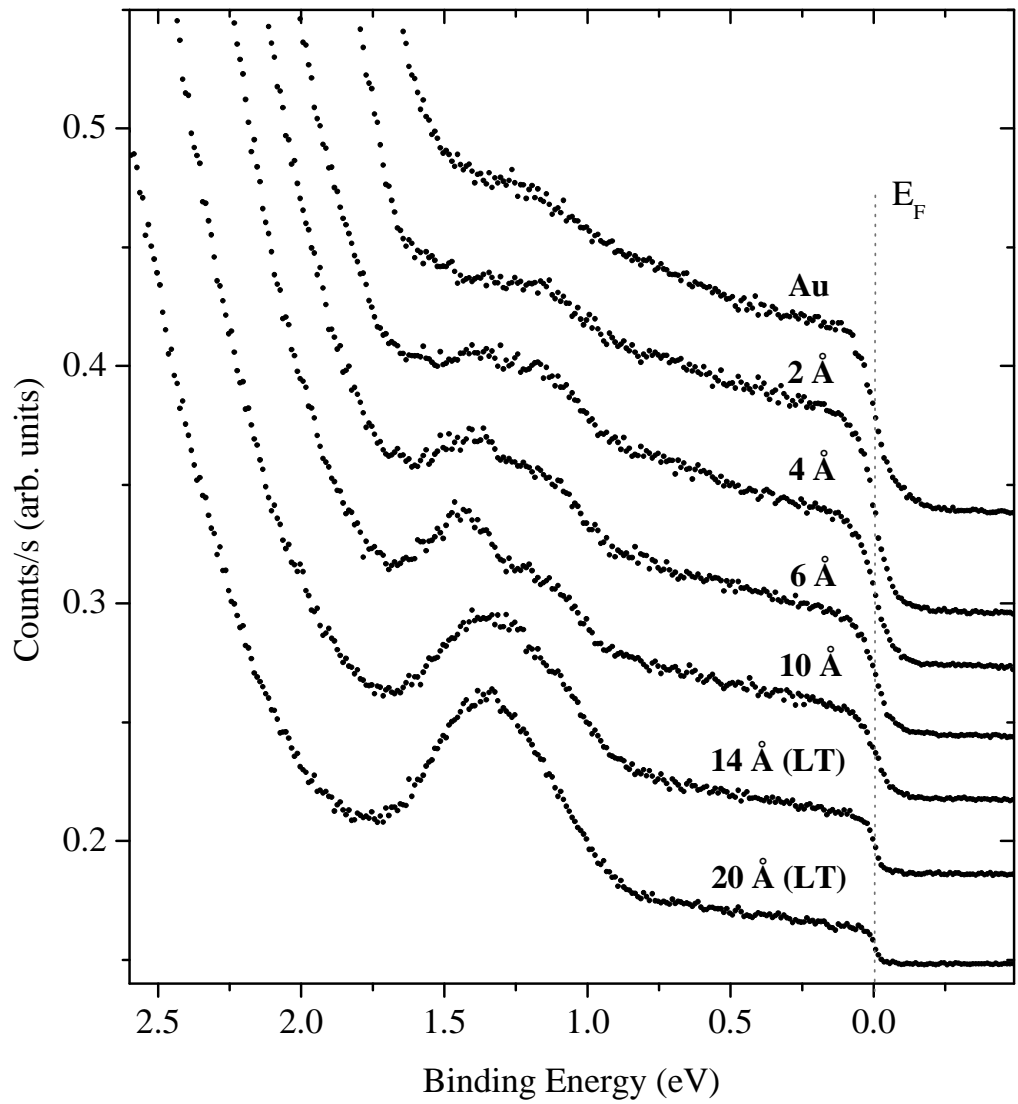

Fig.3 Evangelista et al.

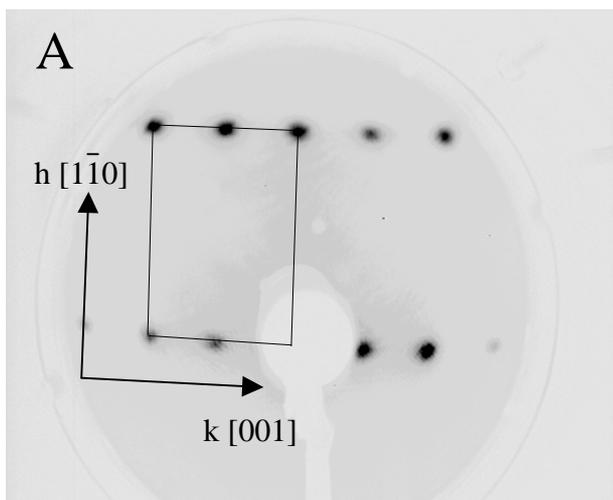

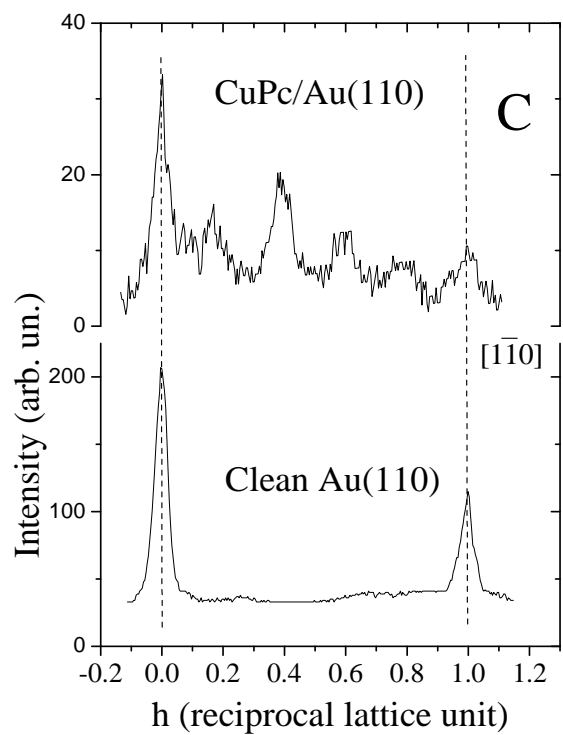

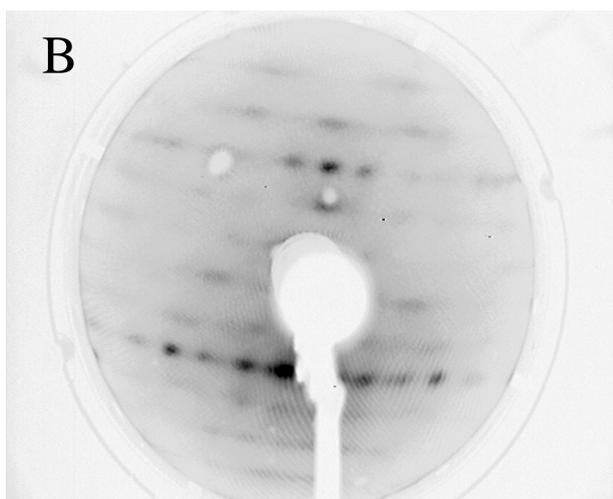

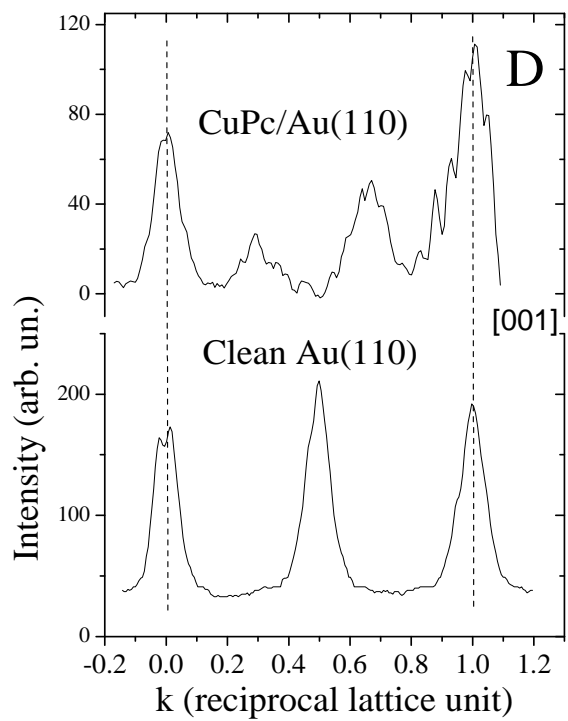

Fig.4 Evangelista et al.